\documentclass[letterpaper, 10 pt, conference]{ieeeconf}  % Comment this line out if you need a4paper

\IEEEoverridecommandlockouts                              % This command is only needed if
\usepackage{graphics,graphicx} % for pdf, bitmapped graphics files
\usepackage{epsfig} % for postscript graphics files
\usepackage{mathptmx} % assumes new font selection scheme installed
\usepackage{times} % assumes new font selection scheme installed
\usepackage{amsmath} % assumes amsmath package installed
\usepackage{amssymb}  % assumes amsmath package installed
\usepackage{dcolumn}% Align table columns on decimal point
\usepackage{bm}% bold math
\usepackage{hyperref}% add hypertext capabilities

\renewcommand{\baselinestretch}{1.1}
\newtheorem{theorem}{Theorem}[section]
\newtheorem{corollary}{Corollary}[section]

\newtheorem{lemma}{Lemma}[section]
\newtheorem{definition}{Definition}

\newtheorem{Proposition}{Proposition}[section]
\newtheorem{Example}{Example}[section]
\newtheorem{Assumption}{Assumption}[section]
\newtheorem{Algorithm}{Algorithm}[section]
\newtheorem{Remark}{Remark}[section]

\renewcommand{\theequation}{\arabic{section}.\arabic{equation}}
\renewcommand{\thesection}{\arabic{section}}

\def\be{\begin{Example}}
\def\ee{\end{Example}}
\def\bt{\begin{theorem}}
\def\et{\end{theorem}\vskip 6pt}
\def\bl{\begin{lemma}}
\def\el{\end{lemma}}
\def\ep{\end{Proposition}}
\def\bp{\begin{Proposition}}
\def\bd{\begin{definition}}
\def\ed{\end{definition}}
\def\ba{\begin{Algorithm}}
\def\ea{\end{Algorithm}}
\def\bs{\begin{Assumption}}
\def\es{\end{Assumption}}
\def\br{\begin{Remark}}
\def\er{\end{Remark}}

\title{\LARGE \bf
Quantum filtering for multiple measurements driven by fields in single-photon states*
}

\author{Zhiyuan Dong$^{1}$ Guofeng Zhang$^{1}$ and Nina H. Amini$^{3}$% <-this % stops a space
\thanks{*This work was financially supported in part by the Hong Kong Research Grant Council (RGC).}% <-this % stops a space
\thanks{$^{1}$Zhiyuan Dong and Guofeng Zhang are with the Department of Applied Mathematics, The Hong Kong Polytechnic University,
        Hung Hom Hong Kong, China
        {\tt\small zhiyuan.dong@connect.polyu.hk, guofeng.zhang@polyu.edu.hk}}%
\thanks{$^{3}$Nina H. Amini is with CNRS, Laboratoire des signaux et syst\`{e}mes (L2S), CentraleSup\'{e}lec, 3 rue Joliot Curie, 91192 Gif-Sur-Yvette, France
        {\tt\small nina.amini@lss.supelec.fr}}%
}

\begin{document}

\maketitle
\thispagestyle{empty}
\pagestyle{empty}

%%%%%%%%%%%%%%%%%%%%%%%%%%%%%%%%%%%%%%%%%%%%%%%%%%%%%%%%%%%%%%%%%%%%%%%%%%%%%%%%
\begin{abstract}

In this paper, we derive the stochastic master equations for quantum systems driven by a single-photon input state which is contaminated by quantum vacuum noise.  To improve estimation performance, quantum filters based on multiple-channel measurements are designed.  Two cases, namely diffusive plus Poissonian measurements and two diffusive measurements, are considered.

\end{abstract}

%%%%%%%%%%%%%%%%%%%%%%%%%%%%%%%%%%%%%%%%%%%%%%%%%%%%%%%%%%%%%%%%%%%%%%%%%%%%%%%%
\section{INTRODUCTION}

When light interacts with a quantum system, e.g., a two-level atom or an optical cavity, partial system information can be transferred to the output light. The output light may be measured, say via homodyne detection, to produce photocurrent upon which the state of quantum system can be conditioned. The stochastic evolution of the conditional system state is usually called quantum trajectory. Quantum filter can be designed to estimate these trajectories \cite{belavkin1989nondemolition, belavkin1995quantum, gardiner2004quantum, GOUGH12QUANTUM, SONG13MULTI, wiseman2009quantum}.

In quantum optics, the quantum filtering problem is known under the names of stochastic master equation and quantum trajectory theory \cite{carmichael2009open, gardiner2004quantum, wiseman2009quantum}. It was first developed by Belavkin within a framework of continuous non-demolition quantum measurement \cite{belavkin1989nondemolition, belavkin1995quantum}. The formalism of quantum filtering for Gaussian input fields, including the vacuum state, coherent state, squeezed state and thermal state, have been considered and well studied \cite{dum1992monte, gardiner2004quantum, Hendra14QUANTUM, wiseman2009quantum}. With a variety of experimental architectures, such as cavity quantum electrodynamics (QED) \cite{mckeever2004deterministic}, circuit QED \cite{eichler2011experimental} and quantum dots in semiconductors \cite{yuan2002electrically}, nonclassical states of light have also been discussed in connection with quantum networks. A range of nonclassical states, single-photon states and coherent states have been considered in \cite{GOUGH12QUANTUM}. Particularly, the master equations and stochastic master equations are presented for an arbitrary quantum system probed by a continuous-mode single-photon input field. As an application, the conditional dynamics for the cross phase modulation in a doubly resonant cavity are considered in \cite{carvalho2012cavity}, where both homodyne detection and photon-counting measurements are simulated for a cavity driven by a single-photon input field. The interaction of a two-level atom with a propagating mode single-photon in free space has been discussed in the literature, see e.g., \cite{wang2011efficient}. The dependence of the atomic excitation probability on the temporal and spectral features of single-photon pulse shapes and coherent states pulse shapes are also considered in \cite{wang2011efficient}.

In real physical experiments, there may be limitations for the case of single measurements due to the existence of noise. To circumvent this imperfection, quantum filtering problem with multiple output fields has been developed using quantum trajectory theory with multi-input-multi-output (MIMO) quantum feedback \cite{chia2011quantum}.  A finite dimensional discrete-time Markov model in the cases of perfect and imperfect measurements are described in \cite{amini2013feedback}. For the state estimations used in the feedback scheme, the quantum filters are discussed and a general robustness property for perfect and imperfect measurements are proved. An experimental implementation has been conducted by using the photon box and closed-loop simulations are also presented~\cite{Sayrin2011}. The observed system in \cite{amini2014stability} is assumed to be governed by a continuous-time stochastic master equation driven by Wiener and Poisson processes. Particularly, the incompleteness and errors in measurements have been taken into account and the measurement imperfections are modeled by a stochastic matrix.

In this paper, we extend the single-photon filtering framework proposed in \cite{carvalho2012cavity, GOUGH12QUANTUM} by including imperfect measurements. More specifically, we study the case when the output light field is corrupted by a vacuum noise. We show how to design filters based on multiple measurements to achieve desired estimation performance. Two scenarios are studied, 1) homodyne plus homodyne detection and 2) homodyne plus photon-counting detection.

The paper is organized as follows. In section~\ref{preliminary}, we review some basic theory such as open quantum systems, series products and quantum filtering. In section~\ref{section3}, we propose quantum filters for the cases of joint homodyne detection and photon-counting measurements and both homodyne detection measurements respectively. We conclude this paper in Section \ref{sec:conc}.

\emph{Notation}. Let $i$ be the imaginary unit and $|0\rangle$ be the vacuum state of the free field. Serif symbols are used for Hilbert spaces, e.g. \textsf{H}. The Hilbert space adjoint or complex conjugate is indicated by $\ast$. The complex conjugate transpose will be denoted by $\dag$, i.e. $X^\dag=(X^\ast)^T$. We will use $\ast$ and $\dag$ interchangeably for single-element operators. The inner product of $X$ and $Y$ in Hilbert space is given by $\langle X,Y\rangle$. The commutator is defined to be $[A,B]=AB-BA$. We set $\mathcal{D}_AB\equiv A^{\dag}BA-\frac{1}{2}(A^{\dag}AB+BA^{\dag}A)$ and $\mathcal{D}^\star_AB\equiv ABA^{\dag}-\frac{1}{2}(A^{\dag}AB+BA^{\dag}A)$. With the description of an operator-valued parametrization of a system, $G=(S,L,H)$, the associated superoperators are
\begin{equation}\nonumber\begin{aligned}
{\rm Lindbladian}:~&\mathcal{L}_GX\equiv-i[X,H]+\mathcal{D}_LX,\\
{\rm Liouvillian}:~&\mathcal{L}^\star_G\rho\equiv-i[X,\rho]+\mathcal{D}^\star_L\rho,
\end{aligned}\end{equation}
and for traceclass $\rho$ and bounded $X$, $\mathrm{Tr}\{\rho\mathcal{L}_GX\}=\mathrm{Tr}\{X\mathcal{L}^\star_G\rho\}$. Finally, $\otimes$ denotes the tensor product.

\section{\label{preliminary}PRELIMINARY}

\subsection{Open Quantum Systems}

The system model we discuss is an arbitrary quantum system $G$ driven by a single-photon input field. Here, we will describe the system by using the triple language $(S,L,H)$ \cite{GOUGH09SERIES,ZJ12}. The scattering operator $S$ is unitary, which satisfies $S^{\dag}S=SS^{\dag}=I$. The coupling between system and field is described by the operator $L$ and the self-adjoint operator $H$ is the initial Hamiltonian of the system.

The input field is represented by annihilation operator $b(t)$ and creation operator $b^\dag(t)$ on the Fock space $\textsf{H}_F$, which satisfies $[b(t),b^\dag(s)]=\delta(t-s)$. The integrated annihilation and creation operators, together with the gauge process are given by
\begin{equation}\nonumber\begin{aligned}
B(t)&=\int^t_0b(s)ds,~~B^\dag(t)=\int^t_0b^\dag(s)ds,\\
\Lambda(t)&=\int^t_0b^\dag(s)b(s)ds.
\end{aligned}\end{equation}

In this paper, we assume that these quantum stochastic processes are canonical, that is, their products satisfy the following It$\bar{\mathrm{o}}$ table
\begin{equation}\begin{aligned}
\begin{array}{c|cccc}
  \times & dt & dB & d\Lambda & dB^\dag \\ \hline
  dt & 0 & 0 & 0 & 0 \\
  dB & 0 & 0 & dB & dt \\
  d\Lambda & 0 & 0 & d\Lambda & dB^\dag \\
  dB^\dag & 0 & 0 & 0 & 0
\end{array}.
\end{aligned}\end{equation}

The dynamical evolution can be described by a unitary operator $U(t)$ on the tensor product Hilbert space $\textsf{H}_S\otimes\textsf{H}_F$ which is given by the following quantum stochastic differential equations (QSDE)
\begin{equation}\begin{aligned}
dU(t)=&\bigg\{(S-I)d\Lambda(t)+LdB^\dag(t)\\
&-L^{\dag}SdB(t)-\left(\frac{1}{2}L^{\dag}L+iH\right)dt\bigg\}U(t)
\end{aligned}\end{equation}
where $U(0)=I$.

In Heisenberg picture, the system operator $X$ is given by a joint operator $j_t(X)=U^{\dag}(t)(X\otimes I_{\mathrm{field}})U(t)$ on $\textsf{H}_S\otimes\textsf{H}_F$. By the quantum It$\bar{\mathrm{o}}$ product rule and table, the temporal evolution of $j_t(X)=X(t)$ is derived as
\begin{equation}\begin{aligned}
dj_t(X)=&j_t(\mathcal{L}_GX)dt+j_t([L^\dag,X]S)dB(t)\\
&+j_t(S^\dag[X,L])dB^{\dag}(t)+j_t(S^{\dag}XS-X)d\Lambda(t).
\end{aligned}\end{equation}
The output fields are defined by
\begin{equation}\nonumber\begin{aligned}
B_{\mathrm{out}}(t)&=U^\dag(t)(I_{\mathrm{system}}\otimes B(t))U(t),\\
\Lambda_{\mathrm{out}}(t)&=U^\dag(t)(I_{\mathrm{system}}\otimes\Lambda(t))U(t),
\end{aligned}\end{equation}
and by It$\bar{\mathrm{o}}$ calculus, we can find the following evolution
\begin{equation}\label{outputevol}\begin{aligned}
dB_{\mathrm{out}}(t)=&S(t)dB(t)+L(t)dt,\\
d\Lambda_{\mathrm{out}}(t)=&S^\ast(t)d\Lambda(t)S^T(t)+S^\ast(t)dB^\ast(t)L^T(t)\\
&+L^\ast(t)dB^T(t)S^T(t)+L^\ast(t)L^T(t)dt.
\end{aligned}\end{equation}

Homodyne and photon-counting detections are the most commonly used measurement methods in quantum filtering \cite{SONG13MULTI}. By using homodyne detection, the measurement is given by quadrature phase
\begin{equation}\begin{aligned}
Y(t)=U^\dag(t)(I_{\mathrm{system}}\otimes(B(t)+B^\dag(t)))U(t),
\end{aligned}\end{equation}
while in the photon-counting case
\begin{equation}\begin{aligned}
Y(t)=\Lambda_{\mathrm{out}}(t)=U^\dag(t)(I_{\mathrm{system}}\otimes\Lambda(t))U(t).
\end{aligned}\end{equation}
Both of the measurements satisfy the following commutation relations $\left[Y(s),Y(t)\right]=0$, $0\leq s\leq t$.
%\begin{equation}\begin{aligned}
%\left[Y(s),Y(t)\right]=0,~~0\leq s\leq t.
%\end{aligned}\end{equation}

\subsection{The Concatenation and Series Products}

\subsubsection{Concatenation Product \cite{GOUGH09SERIES}}

\begin{figure}
\centering
\includegraphics[scale=0.3]{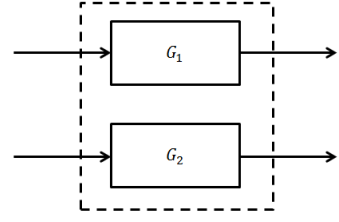}
\caption{Concatenation product.}
\label{Concatenation product}
\end{figure}

Given two systems $G_1=(S_1,L_1,H_1)$ and $G_2=(S_2,L_2,H_2)$, see Fig.~\ref{Concatenation product}, we define the concatenation product to be the system $G_1\boxplus G_2$ by
\begin{equation}\begin{aligned}
G_1\boxplus G_2=\left(\left[
                         \begin{array}{cc}
                           S_1 & 0 \\
                           0 & S_2 \\
                         \end{array}
                       \right]
,\left[
                          \begin{array}{c}
                            L_1 \\
                            L_2 \\
                          \end{array}
                        \right]
,H_1+H_2\right).
\end{aligned}\end{equation}

\subsubsection{Series Product \cite{GOUGH09SERIES}}

\begin{figure}
\centering
\includegraphics[scale=0.3]{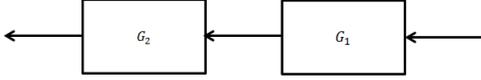}
\caption{Series product.}
\label{Series product}
\end{figure}

Given two systems $G_1=(S_1,L_1,H_1)$ and $G_2=(S_2,L_2,H_2)$ with the same number of field channels, see Fig.~\ref{Series product}, we define the series product $G_2\vartriangleleft G_1$ by $G_2\vartriangleleft G_1=\left(S_2S_1,L_2+S_2L_1,H_1+H_2+\mathrm{Im}\{L^\dag_2S_2L_1\}\right)$.
%\begin{equation}\begin{aligned}
%G_2\vartriangleleft G_1=\left(S_2S_1,L_2+S_2L_1,H_1+H_2+\mathrm{Im}\{L^\dag_2S_2L_1\}\right).
%\end{aligned}\end{equation}

\subsection{Quantum Filtering}

The quantum conditional expectation is defined by
\begin{equation}\begin{aligned}
\hat{X}(t)=\pi_t(X)=\mathbb{E}[j_t(X)|\mathcal{Y}_t],
\end{aligned}\end{equation}
where $\mathcal{Y}_t$ is generated by $\{Y(s):0\leq s\leq t\}$. Generally the quantum filtering problem is about minimizing the least mean-squares estimate $\mathbb{E}[\{\hat{X}(t)-j_t(X)\}^2]$ of system observables $j_t(X)$ based on the past measurement information $\mathcal{Y}_t$. Furthermore, we note that the set of observables $\{Y(s):0\leq s\leq t\}$ is self-commuting $[Y(t),Y(s)]=0$, $s\leq t$.
%\begin{equation}\begin{aligned}
%\left[Y(t),Y(s)\right]=0,~~s\leq t.
%\end{aligned}\end{equation}
The quantum conditional expectation is well-defined since it satisfies the non-demolition property $[X(t),Y(s)]=0$, $s\leq t$.
%\begin{equation}\begin{aligned}
%\left[X(t),Y(s)\right]=0,~~s\leq t.
%\end{aligned}\end{equation}

\section{\label{section3}QUANTUM FILTER FOR MULTIPLE MEASUREMENTS}

\begin{figure}
\centering
\includegraphics[scale=0.3]{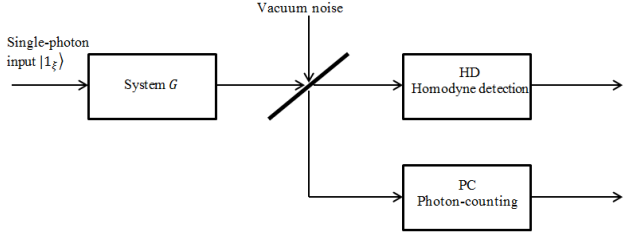}
\caption{Simultaneous homodyne detection and photon-counting at the outputs of a beam splitter in quantum system.}
\label{Simultaneous}
\end{figure}

In this section, we mainly present the stochastic master equations for a quantum system interacting with a single-photon state, see Fig.~\ref{Simultaneous}. The single-photon state is defined in subsection~\ref{Single-photon state}, the extended system is briefly reviewed and the relation between expectation for system and for extended system is discussed in subsection~\ref{extended}, the general quantum filter for multiple compatible measurements is introduced in subsection~\ref{vacuum input}. For the single-photon input state, we derive the filtering equations in subsection~\ref{filter single-photon} and~\ref{bothhd}. An illustrating example is given in subsection \ref{susec:simulation}.

\subsection{\label{Single-photon state}Continuous-mode Single-photon State}

The creation operator for a photon with wave packet $\xi(t)$ in time domain is defined as
\begin{equation}\begin{aligned}
B^\ast(\xi)=\int^{\infty}_0\xi(t)b^\ast(t)dt,
\end{aligned}\end{equation}
with the normalization condition $\int^\infty_0|\xi(t)|^2dt=1$. Then the single-photon state is given by
\begin{equation}\begin{aligned}
|1_{\xi}\rangle=B^\ast(\xi)|0\rangle.
\end{aligned}\end{equation}

%Similarly, we can define the single-photon state in frequency domain
%\begin{equation}\begin{aligned}
%|1_{\xi}\rangle=\int^{\infty}_{-\infty}\hat{\xi}(\omega)\hat{b}^\ast(\omega)d\omega|0\rangle,
%\end{aligned}\end{equation}
%where $\hat{\xi}(\omega)$ is the Fourier transform of $\xi(t)$.

The original departure of the master and filter equations between single-photon input and the vacuum case are given by the following identities
\begin{equation}\begin{aligned}
dB(t)|1_\xi\rangle&=\xi(t)dt|0\rangle,\\
d\Lambda(t)|1_\xi\rangle&=\xi(t)dB^\dag(t)|0\rangle.
\end{aligned}\end{equation}

\subsection{The Extended System}\label{extended}

%There are some methods for generating a single-photon state. Gheri, \emph{et al}. \cite{gheri1998photon} proposed a cavity initially with one photon as the auxiliary emitter while Gough, \emph{et al}. \cite{GOUGH12QUANTUM,gough2013quantum} gave a two-level system initially in the excited state $|\uparrow\rangle$. In this section, we will use the latter and cascade approach.

We construct a quantum signal generating filter $M=(S_M,L_M,H_M)$, which is usually called ancilla. Cascading this ancilla system model $M$ with the quantum system $G$, then we get the extended system $G_T=G\vartriangleleft M$. Since the extended system $G_T$ is driven by vacuum input, the master equation and quantum filter follow from the known result in subsection~\ref{vacuum input} with the parameters for the extended system $G_T$ \cite{GOUGH12QUANTUM}. The interaction with the vacuum input is given by
\begin{equation}\begin{aligned}
(S_M,L_M,H_M)=(I,\lambda(t)\sigma_-,0),
\end{aligned}\end{equation}
where $\sigma_-$ is the lowing operator from the upper state $|\uparrow\rangle$ to the ground state $|\downarrow\rangle$. It means that the atom decays into its ground state at some stage, creating a single photon in the output. The signal model will output the desired single-photon state $|1_\xi\rangle$ since we can choose the coupling strength $\lambda(t)$
\begin{equation}\begin{aligned}
\lambda(t)=\frac{\xi(t)}{\sqrt{w(t)}},
\end{aligned}\end{equation}
where $w(t)=\int^{\infty}_t|\xi(s)|^2ds$.

By using the cascade connection formalism, we have the extended system $G_T$
\begin{equation}\begin{aligned}
G_T=\left(S,L+\lambda(t)S\sigma_-,H+\lambda(t)\mathrm{Im}\{L^{\dag}S\sigma_-\}\right).
\end{aligned}\end{equation}

Let $\tilde{U}(t)$ be the unitary operator for the joint ancilla, system and field spaces. The following equality can be shown (see~\cite{GOUGH12QUANTUM} for more details)
\begin{equation}\begin{aligned}
\mathbb{E}_{\eta\xi}[X(t)]=\mathbb{E}_{\uparrow\eta0}[\tilde{U}^\dag(t)(I\otimes X\otimes I)\tilde{U}(t)],
\end{aligned}\end{equation}
with initial state $|\uparrow\rangle\otimes|\eta\rangle\otimes|0\rangle$ for arbitrary operator $X(t)$ of the system $G$.

\subsection{\label{vacuum input}Quantum Filter for Multiple Measurements Driven by Vacuum Input}

To derive the quantum filter for system driven by a single-photon input state, we firstly introduce the result of multiple measurements with vacuum state input.

\begin{lemma}\label{peterlemma} (\cite[Theorem 3.2]{woolley2015quantum})
Let $\{Y_{i,t},i=1,2,\ldots,N\}$ be a set of $N$ compatible measurement outputs for a quantum system $G$. With vacuum initial state, the corresponding joint measurement quantum filter is given by
\begin{equation}\begin{aligned}
d\hat{X}=\pi_t[\mathcal{L}_G(X_t)]dt+\displaystyle{\sum^N_{i=1}}\beta_{i,t}dW_{i,t},
\end{aligned}\end{equation}
where $dW_{i,t}=dY_{i,t}-\pi_t(dY_{i,t})$ is a martingale process for each measurement output and $\beta_{i,t}$ is the corresponding gain given by
\begin{equation}\label{beta12}\begin{aligned}
\zeta^T&=\pi_t(X_tdY^T_t)-\pi_t(X_t)\pi_t(dY^T_t)+\pi_t\left([L^\dag_t,X_t]S_tdBdY^T_t\right),\\
\Sigma&=\pi_t(dY_tdY^T_t), ~~ \beta=\Sigma^{-1}\zeta,
\end{aligned}\end{equation}
where $\Sigma$ is assumed to be non-singular.
\end{lemma}

\begin{Remark}
A general measurement equation, which is a function of annihilation, creation and conservation processes in the output field, is defined as \cite{woolley2015quantum}
\begin{equation}\label{fg}\begin{aligned}
dY(t)=F^\ast dB^\ast_{\mathrm{out}}(t)+FdB_{\mathrm{out}}(t)+G\mathrm{diag}(d\Lambda_{\mathrm{out}}(t)).
\end{aligned}\end{equation}
Particularly, a combination of homodyne detection and photon-counting measurement is given by
\begin{equation}\nonumber\begin{aligned}
F=\left[
    \begin{array}{cc}
      1 & 0 \\
      0 & 0 \\
    \end{array}
  \right], G=\left[
               \begin{array}{cc}
                 0 & 0 \\
                 0 & 1 \\
               \end{array}
             \right].
\end{aligned}\end{equation}
\end{Remark}

\subsection{\label{filter single-photon}Quantum Filter for Joint Homodyne and Photon-counting Detections Driven by Single-photon State}

Suppose that the system is in an initial state $\rho_0=|\eta\rangle\langle\eta|$ and the single-photon input state is $|1_{\xi}\rangle$. For a given system operator $X$, we define the expectation
$$\omega^{jk}_t(X)=\mathbb{E}_{jk}[j_t(X)]=\langle\eta\phi_j|X|\eta\phi_k\rangle,~~j,k=0,1,$$
where $\phi_j=\left\{\begin{array}{ll}
                      |0\rangle, & j=0; \\
                      |1_{\xi}\rangle, & j=1.
                    \end{array}
\right.$

The quantum filter for the conditional expectation for the system $G$ driven by a single-photon field is given by
\begin{equation}\begin{aligned}
\pi^{11}_t(X)=\mathbb{E}_{\eta\xi}[X(t)|Y(s),0\leq s\leq t],
\end{aligned}\end{equation}
and the quantum filter for the extended system $G_T=G\vartriangleleft M$ driven by vacuum input is defined as
\begin{equation}\begin{aligned}
\tilde{\pi}_t(A\otimes X)=\mathbb{E}_{\uparrow\eta0}[\tilde{U}^\dag(t)(A\otimes X)\tilde{U}(t)|I\otimes Y(s),0\leq s\leq t],
\end{aligned}\end{equation}
where $A$ is an ancilla operator and $X$ is a system operator.

\begin{figure}
\centering
\includegraphics[scale=0.27]{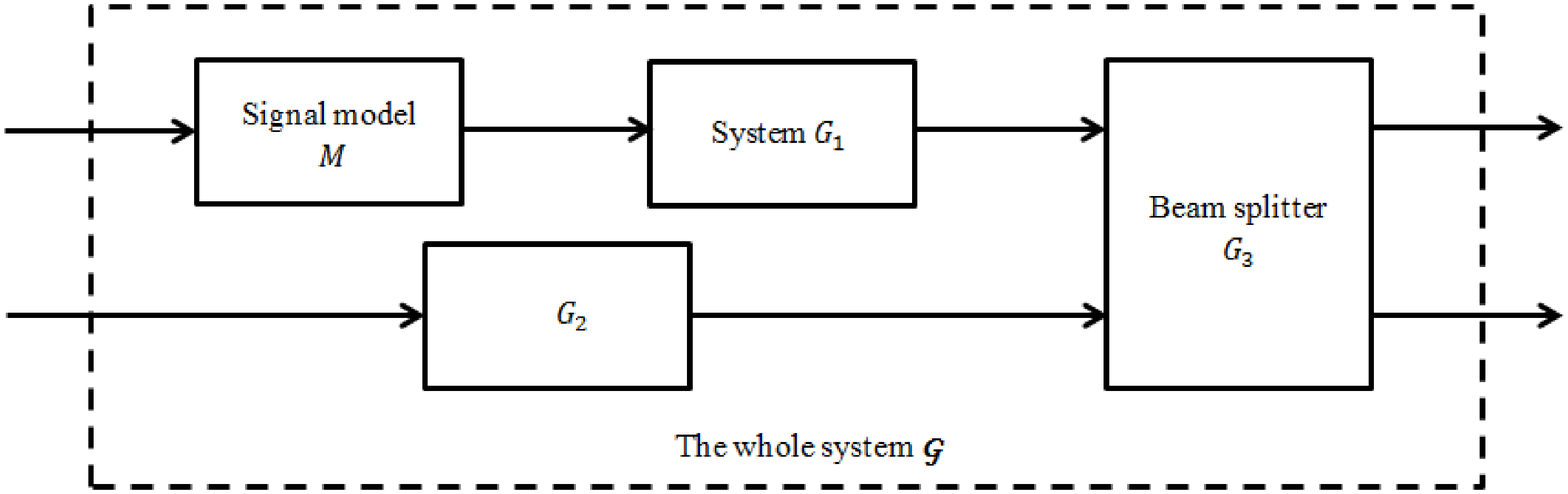}
\caption{Quantum system depiction of Figure 3.}
\label{Depiction}
\end{figure}

The whole system $\mathcal{G}$ with the measurements in Fig.~\ref{Simultaneous} can be depicted as shown in Fig.~\ref{Depiction}. $G_1=(S,L,H)$ is the original system $G$, which has been connected with a signal model (ancilla) $M=(I,\lambda(t)\sigma_-,0)$. By introducing a second open quantum system $G_2=(1,0,0)$, we concatenate the vacuum noise into our system. The last open quantum system is a beam splitter $G_3=(S_b,0,0)$, where
\begin{equation}\begin{aligned}
S_b=\left[
       \begin{array}{cc}
         \sqrt{1-r^2}e^{i\theta} & re^{i(\theta+\frac{\pi}{2})} \\
         re^{i(\theta+\frac{\pi}{2})} & \sqrt{1-r^2}e^{i\theta} \\
       \end{array}
     \right],~~0 \leq r \leq 1.
\end{aligned}\end{equation}

By the concatenation and series product, the whole system $\mathcal{G}$ is given by
\begin{equation}\begin{aligned}
\mathcal{G}=G_3\vartriangleleft[(G_1\vartriangleleft M)\boxplus G_2]=(S_t,L_t,H_t),
\end{aligned}\end{equation}
where $S_t=S_b\left[
                \begin{array}{cc}
                  S & 0 \\
                  0 & 1 \\
                \end{array}
              \right]
$, $L_t=\left[
          \begin{array}{c}
            L+\lambda(t)S\sigma_- \\
            0 \\
          \end{array}
        \right]
$, $H_t=H+\lambda(t)\mathrm{Im}\{L^{\dag}S\sigma_-\}$.

Furthermore, the Lindblad superoperator $\mathcal{L}_{\mathcal{G}}(A\otimes X)$ for the whole system $\mathcal{G}$ can be expressed in the following form
\begin{equation}\label{supero}\begin{aligned}
\mathcal{L}_{\mathcal{G}}(A\otimes X)=&A\otimes\mathcal{L}_{G}X+(\mathcal{D}_{L_M}A)\otimes X+
L^\dag_MA\otimes S^\dag[X,L]\\
&+AL_M\otimes[L^\dag,X]S+L^\dag_MAL_M\otimes(S^{\dag}XS-X),
\end{aligned}\end{equation}
where $A$ is any operator of ancilla and $X$ is the system operator.

In what follows, we denote $B_{i,t}$, which is a vacuum state, is the input of signal model $M$ and $B_{v,t}$ is the vacuum noise for system $G_2$, then the total input, together with gauge process for the whole system $\mathcal{G}$ are given by
\begin{equation}\nonumber\begin{aligned}
B_t=\left[
      \begin{array}{c}
        B_{i,t} \\
        B_{v,t} \\
      \end{array}
    \right], \Lambda_t=\left[
                         \begin{array}{cc}
                           \Lambda_{i,t} & \Lambda_{iv,t} \\
                           \Lambda_{vi,t} & \Lambda_{v,t} \\
                         \end{array}
                       \right].
\end{aligned}\end{equation}

By the evolution of output fields \eqref{outputevol}, the measurements stochastic equations are given by
\begin{equation}\label{dy1}\begin{aligned}
dY_{1,t}=&\sqrt{1-r^2}\Big\{\left[e^{i\theta}(L+SL_M)+e^{-i\theta}(L^\dag+L^\dag_MS^\dag)\right]dt\\
&+e^{i\theta}SdB_{i,t}+e^{-i\theta}S^{\dag}dB^\dag_{i,t}\Big\}\\
&+ir\left(e^{i\theta}dB_{v,t}-e^{-i\theta}dB^\dag_{v,t}\right),
\end{aligned}\end{equation}
and
\begin{equation}\begin{aligned}
dY_{2,t}=&r^2\Big[Sd\Lambda_{i,t}S^\dag+(L+SL_M)S^{\dag}dB^\dag_{i,t}+S(L^\dag+L^\dag_MS^\dag)dB_{i,t}\\
&+(L^\dag+L^\dag_MS^\dag)(L+SL_M)dt\Big]+(1-r^2)d\Lambda_{v,t}\\
&+ir\sqrt{1-r^2}\Big[Sd\Lambda_{vi,t}-S^{\dag}d\Lambda_{iv,t}+(L+SL_M)dB^\dag_{v,t}\\
&-(L^\dag+L^\dag_MS^\dag)dB_{v,t}\Big],
\end{aligned}\end{equation}
where $dY_{1,t}$ is the first channel with homedyne detection and $dY_{2,t}$ is the second channel with photon-counting measurement. Then the expectation and correlation of the measurement can be derived as
\begin{equation}\begin{aligned}
&\tilde{\pi}_t(dY_{1,t})=\sqrt{1-r^2}\tilde{\pi}_t\left[e^{i\theta}(L+SL_M)+e^{-i\theta}(L^\dag+L^\dag_MS^\dag)\right]dt,\\
&\tilde{\pi}_t(dY_{2,t})=\tilde{\pi}_t(dY_{2,t}dY_{2,t})=r^2\tilde{\pi}_t\left[(L^\dag+L^\dag_MS^\dag)(L+SL_M)\right]dt,\\
&\tilde{\pi}_t(dY_{1,t}dY_{1,t})=dt,\\
&\tilde{\pi}_t(dY_{1,t}dY_{2,t})=\tilde{\pi}_t(dY_{2,t}dY_{1,t})=0.\\
\end{aligned}\end{equation}

Thus, the corresponding gain $\beta=[\beta_1 ~ \beta_2]$ can be calculated by \eqref{beta12}
\begin{equation}\label{beta1}\begin{aligned}
\beta_1=&\sqrt{1-r^2}e^{i\theta}\tilde{\pi}_t\big(A\otimes XL+AL_M\otimes XS\big)\\
&+\sqrt{1-r^2}e^{-i\theta}\tilde{\pi}_t\big(A\otimes L^{\dag}X+L^\dag_MA\otimes S^{\dag}X\big)\\
&-\sqrt{1-r^2}\tilde{\pi}_t\big(A\otimes X\big)\\ &\times\tilde{\pi}_t\left[e^{i\theta}(L+SL_M)+e^{-i\theta}(L^\dag+L^\dag_MS^\dag)\right],\\
\end{aligned}\end{equation}
\begin{equation}\begin{aligned}
\beta_2=&\big[\tilde{\pi}_t(L^{\dag}L+L^\dag_MS^{\dag}L+L^{\dag}SL_M+L^\dag_ML_M)\big]^{-1}\\
&\times\tilde{\pi}_t\big(A\otimes L^{\dag}XL+L^\dag_MA\otimes S^{\dag}XL+AL_M\otimes L^{\dag}XS\\
&+L^\dag_MAL_M\otimes S^{\dag}XS\big)-\tilde{\pi}_t\big(A\otimes X\big),
\end{aligned}\end{equation}
where $A$ is any ancilla operator and $X$ is the system operator.

Finally, by Lemma~\ref{peterlemma}, the joint measurement quantum filter for the whole system $\mathcal{G}$ driven by vacuum input state is given by
\begin{equation}\begin{aligned}
d\tilde{\pi}_t(A\otimes X)=&\tilde{\pi}_t(\mathcal{L}_{\mathcal{G}}(A\otimes X))dt+
\beta_1\left[dY_{1,t}-\tilde{\pi}_t(dY_{1,t})\right]\\
&+\beta_2\left[dY_{2,t}-\tilde{\pi}_t(dY_{2,t})\right],
\end{aligned}\end{equation}
where the Lindblad superoperator $\mathcal{L}_{\mathcal{G}}(A\otimes X)$ is defined in \eqref{supero}.

If we define \cite{GOUGH12QUANTUM}
\begin{equation}\begin{aligned}
\pi^{jk}_t(X)=\frac{\tilde{\pi}_t(Q_{jk}{\otimes}X)}{w_{jk}},~~j,k=0,1,
\end{aligned}\end{equation}
where $Q_{jk}$ and $w_{jk}$ are given by
\begin{equation}\nonumber\begin{aligned}
&Q_{jk}=\left[
         \begin{array}{cc}
           Q_{00} & Q_{01} \\
           Q_{10} & Q_{11} \\
         \end{array}
       \right]=\left[
                 \begin{array}{cc}
                   \sigma_+\sigma_- & \sigma_+ \\
                   \sigma_- & I \\
                 \end{array}
               \right],\\
                &w_{jk}=\left[
                                 \begin{array}{cc}
                                   w_{00} & w_{01} \\
                                   w_{10} & w_{11} \\
                                 \end{array}
                               \right]=\left[
                                         \begin{array}{cc}
                                           w(t) & \sqrt{w(t)} \\
                                           \sqrt{w(t)} & 1 \\
                                         \end{array}
                                       \right],
\end{aligned}\end{equation}
we obtain the following theorem which presents the quantum filter for the system $G$ driven by single-photon state $|1_{\xi}\rangle$.

\begin{theorem}\label{theorem}
Let $\{Y_{i,t},i=1,2\}$ be a combination of homodyne detection and photon-counting measurement for a quantum system $G$. With single-photon input state, the quantum filter for the conditional expectation in the Heisenberg picture is given by (3.36).
\newcounter{mytempeqncnt}
\begin{figure*}[!t]
\normalsize
\setcounter{mytempeqncnt}{\value{equation}}
\setcounter{equation}{35}
\begin{equation}\label{sdehei}\begin{aligned}
d\pi^{11}_t(X)=&\big\{\pi^{11}_t(\mathcal{L}_GX)+\pi^{01}_t(S^\dag[X,L])\xi^\ast(t)+\pi^{10}_t([L^\dag,X]S)\xi(t)+\pi^{00}_t(S^{\dag}XS-X)|\xi(t)|^2\big\}dt\\
&+\sqrt{1-r^2}\big[e^{i\theta}\pi^{11}_t(XL)+e^{-i\theta}\pi^{11}_t(L^{\dag}X)+e^{-i\theta}\pi^{01}_t(S^{\dag}X)\xi^\ast(t)+e^{i\theta}\pi^{10}_t(XS)\xi(t)-\pi^{11}_t(X)K_t\big]dW(t)\\
&+\big\{\nu^{-1}_t\big[\pi^{11}_t(L^{\dag}XL)+\pi^{01}_t(S^{\dag}XL)\xi^\ast(t)+\pi^{10}_t(L^{\dag}XS)\xi(t)+\pi^{00}_t(S^{\dag}XS)|\xi(t)|^2\big]-\pi^{11}_t(X)\big\}dN(t),\\
d\pi^{10}_t(X)=&\left\{\pi^{10}_t(\mathcal{L}_GX)+\pi^{00}_t(S^\dag[X,L])\xi^\ast(t)\right\}dt\\
&+\sqrt{1-r^2}\left[e^{i\theta}\pi^{10}_t(XL)+e^{-i\theta}\pi^{10}_t(L^{\dag}X)+e^{-i\theta}\pi^{00}_t(S^{\dag}X)\xi^\ast(t)-\pi^{10}_t(X)K_t\right]dW(t)\\
&+\left\{\nu^{-1}_t\left[\pi^{10}_t(L^{\dag}XL)+\pi^{00}_t(S^{\dag}XL)\xi^\ast(t)\right]-\pi^{10}_t(X)\right\}dN(t),\\
d\pi^{01}_t(X)=&\left\{\pi^{01}_t(\mathcal{L}_GX)+\pi^{00}_t([L^\dag,X]S)\xi(t)\right\}dt\\
&+\sqrt{1-r^2}\left[e^{i\theta}\pi^{01}_t(XL)+e^{-i\theta}\pi^{01}_t(L^{\dag}X)+e^{i\theta}\pi^{00}_t(XS)\xi(t)-\pi^{01}_t(X)K_t\right]dW(t)\\
&+\left\{\nu^{-1}_t\left[\pi^{01}_t(L^{\dag}XL)+\pi^{00}_t(L^{\dag}XS)\xi(t)\right]-\pi^{01}_t(X)\right\}dN(t),\\
d\pi^{00}_t(X)=&\pi^{00}_t(\mathcal{L}_GX)dt+\sqrt{1-r^2}\left[e^{i\theta}\pi^{00}_t(XL)+e^{-i\theta}\pi^{00}_t(L^{\dag}X)-\pi^{00}_t(X)K_t\right]dW(t)\\
&+\left\{\nu^{-1}_t\left[\pi^{00}_t(L^{\dag}XL)\right]-\pi^{00}_t(X)\right\}dN(t).
\end{aligned}
\end{equation}
\setcounter{equation}{36}
\hrulefill
\vspace*{4pt}
\end{figure*}
%\begin{equation}\label{sdehei}\begin{aligned}
%d\pi^{11}_t(X)=&\big\{\pi^{11}_t(\mathcal{L}_GX)+\pi^{01}_t(S^\dag[X,L])\xi^\ast(t)+\pi^{10}_t([L^\dag,X]S)\xi(t)+\pi^{00}_t(S^{\dag}XS-X)|\xi(t)|^2\big\}dt\\
%&+\sqrt{1-r^2}\big[e^{i\theta}\pi^{11}_t(XL)+e^{-i\theta}\pi^{11}_t(L^{\dag}X)+e^{-i\theta}\pi^{01}_t(S^{\dag}X)\xi^\ast(t)+e^{i\theta}\pi^{10}_t(XS)\xi(t)-\pi^{11}_t(X)K_t\big]dW(t)\\
%&+\big\{\nu^{-1}_t\big[\pi^{11}_t(L^{\dag}XL)+\pi^{01}_t(S^{\dag}XL)\xi^\ast(t)+\pi^{10}_t(L^{\dag}XS)\xi(t)+\pi^{00}_t(S^{\dag}XS)|\xi(t)|^2\big]-\pi^{11}_t(X)\big\}dN(t),\\
%d\pi^{10}_t(X)=&\left\{\pi^{10}_t(\mathcal{L}_GX)+\pi^{00}_t(S^\dag[X,L])\xi^\ast(t)\right\}dt\\
%&+\sqrt{1-r^2}\left[e^{i\theta}\pi^{10}_t(XL)+e^{-i\theta}\pi^{10}_t(L^{\dag}X)+e^{-i\theta}\pi^{00}_t(S^{\dag}X)\xi^\ast(t)-\pi^{10}_t(X)K_t\right]dW(t)\\
%&+\left\{\nu^{-1}_t\left[\pi^{10}_t(L^{\dag}XL)+\pi^{00}_t(S^{\dag}XL)\xi^\ast(t)\right]-\pi^{10}_t(X)\right\}dN(t),\\
%d\pi^{01}_t(X)=&\left\{\pi^{01}_t(\mathcal{L}_GX)+\pi^{00}_t([L^\dag,X]S)\xi(t)\right\}dt\\
%&+\sqrt{1-r^2}\left[e^{i\theta}\pi^{01}_t(XL)+e^{-i\theta}\pi^{01}_t(L^{\dag}X)+e^{i\theta}\pi^{00}_t(XS)\xi(t)-\pi^{01}_t(X)K_t\right]dW(t)\\
%&+\left\{\nu^{-1}_t\left[\pi^{01}_t(L^{\dag}XL)+\pi^{00}_t(L^{\dag}XS)\xi(t)\right]-\pi^{01}_t(X)\right\}dN(t),\\
%d\pi^{00}_t(X)=&\pi^{00}_t(\mathcal{L}_GX)dt+\sqrt{1-r^2}\left[e^{i\theta}\pi^{00}_t(XL)+e^{-i\theta}\pi^{00}_t(L^{\dag}X)-\pi^{00}_t(X)K_t\right]dW(t)\\
%&+\left\{\nu^{-1}_t\left[\pi^{00}_t(L^{\dag}XL)\right]-\pi^{00}_t(X)\right\}dN(t).
%\end{aligned}
%\end{equation}
Here,
\begin{equation}\begin{aligned}
K_t:=&e^{i\theta}\pi^{11}_t(L)+e^{-i\theta}\pi^{11}_t(L^\dag)\\
&+e^{-i\theta}\pi^{01}_t(S^\dag)\xi^\ast(t)+e^{i\theta}\pi^{10}_t(S)\xi(t),\\
\nu_t:=&\pi^{11}_t(L^{\dag}L)+\pi^{01}_t(S^{\dag}L)\xi^\ast(t)\\
&+\pi^{10}_t(L^{\dag}S)\xi(t)+\pi^{00}_t(I)|\xi(t)|^2,
\end{aligned}\end{equation}
the Wiener process $W(t)$ and compensated Poisson process $N(t)$ are given by
\begin{equation}\begin{aligned}
dW(t)=dY_{1,t}-\sqrt{1-r^2}K_tdt,~~dN(t)=dY_{2,t}-r^2\nu_tdt
\end{aligned}\end{equation}
respectively. We have $\pi^{10}_t(X)=\pi^{01}_t(X^\dag)^\dag$, the initial conditions are $\pi^{11}_0(X)=\pi^{00}_0(X)=\langle\eta,X\eta\rangle,~~\pi^{10}_0(X)=\pi^{01}_0(X)=0$.
%\begin{equation}\begin{aligned}
%\pi^{11}_0(X)=\pi^{00}_0(X)=\langle\eta,X\eta\rangle,~~\pi^{10}_0(X)=\pi^{01}_0(X)=0.
%\end{aligned}\end{equation}
\end{theorem}

\begin{Remark}
If we let $r=0$, $\theta=0$, the filter equations reduce to an estimation problem with a single homodyne detection. It can be verified that \eqref{sdehei} would be in the same form in~\cite{GOUGH12QUANTUM}. On the other hand, if we let $r=1$, $\theta=-\frac{\pi}{2}$, the filter equations reduce to an estimation problem with a single photon-counting measurement. It also can be checked that \eqref{sdehei} would reduce to the corresponding case in~\cite{GOUGH12QUANTUM}.
\end{Remark}

If we write $\pi^{jk}_t(X)=\mathrm{Tr}[(\rho^{jk}(t))^{\dag}X]$, by the quantum filter \eqref{sdehei}, we can get the following differential equations for the evolution of $\rho^{jk}(t)$.
\begin{corollary}
With a combination of homodyne detection and photon-counting measurement, the quantum filter for the system $G$ driven by single-photon input state in the Schr$\ddot{\mathrm{o}}$dinger picture is given by (3.40).
\newcounter{mytempeqncntt}
\begin{figure*}[!t]
\normalsize
\setcounter{mytempeqncntt}{\value{equation}}
\setcounter{equation}{39}
\begin{equation}\begin{aligned}
d\rho^{11}(t)=&\left\{\mathcal{L}^\star_G\rho^{11}(t)+[S\rho^{01}(t),L^\dag]\xi(t)+[L,\rho^{10}(t)S^\dag]\xi^\ast(t)+[S\rho^{00}(t)S^\dag-\rho^{00}(t)]|\xi(t)|^2\right\}dt\\
&+\sqrt{1-r^2}\left[e^{-i\theta}\rho^{11}(t)L^\dag+e^{i\theta}L\rho^{11}(t)+e^{i\theta}S\rho^{01}(t)\xi(t)+e^{-i\theta}\rho^{10}(t)S^\dag\xi^\ast(t)-K_t\rho^{11}(t)\right]dW(t)\\
&+\left\{\nu^{-1}_t\left[L\rho^{11}(t)L^\dag+S\rho^{01}(t)L^\dag\xi(t)+L\rho^{10}(t)S^\dag\xi^\ast(t)+S\rho^{00}(t)S^\dag|\xi(t)|^2\right]-\rho^{11}(t)\right\}dN(t),\\
d\rho^{10}(t)=&\left\{\mathcal{L}^\star_G\rho^{10}(t)+[S\rho^{00}(t),L^\dag]\xi(t)\right\}dt\\
&+\sqrt{1-r^2}\left[e^{-i\theta}\rho^{10}(t)L^\dag+e^{i\theta}L\rho^{10}(t)+e^{i\theta}S\rho^{00}(t)\xi(t)-K_t\rho^{10}(t)\right]dW(t)\\
&+\left\{\nu^{-1}_t[L\rho^{10}(t)L^\dag+S\rho^{00}(t)L^\dag\xi(t)]-\rho^{10}(t)\right\}dN(t),\\
d\rho^{01}(t)=&\left\{\mathcal{L}^\star_G\rho^{01}(t)+[L,\rho^{00}(t)S^\dag]\xi^\ast(t)\right\}dt\\
&+\sqrt{1-r^2}\left[e^{-i\theta}\rho^{01}(t)L^\dag+e^{i\theta}L\rho^{01}(t)+e^{-i\theta}\rho^{00}(t)S^\dag\xi^\ast(t)-K_t\rho^{01}(t)\right]dW(t)\\
&+\left\{\nu^{-1}_t[L\rho^{01}(t)L^\dag+L\rho^{00}(t)S^\dag\xi^\ast(t)]-\rho^{01}(t)\right\}dN(t),\\
d\rho^{00}(t)=&\mathcal{L}^\star_G\rho^{00}(t)dt+\sqrt{1-r^2}\left[e^{-i\theta}\rho^{00}(t)L^\dag+e^{i\theta}L\rho^{00}(t)-K_t\rho^{00}(t)\right]dW(t)\\
&+\left\{\nu^{-1}_t[L\rho^{00}(t)L^\dag]-\rho^{00}(t)\right\}dN(t).
\end{aligned}\end{equation}
\setcounter{equation}{40}
\hrulefill
\vspace*{4pt}
\end{figure*}
%\begin{equation}\begin{aligned}
%d\rho^{11}(t)=&\left\{\mathcal{L}^\star_G\rho^{11}(t)+[S\rho^{01}(t),L^\dag]\xi(t)+[L,\rho^{10}(t)S^\dag]\xi^\ast(t)+[S\rho^{00}(t)S^\dag-\rho^{00}(t)]|\xi(t)|^2\right\}dt\\
%&+\sqrt{1-r^2}\left[e^{-i\theta}\rho^{11}(t)L^\dag+e^{i\theta}L\rho^{11}(t)+e^{i\theta}S\rho^{01}(t)\xi(t)+e^{-i\theta}\rho^{10}(t)S^\dag\xi^\ast(t)-K_t\rho^{11}(t)\right]dW(t)\\
%&+\left\{\nu^{-1}_t\left[L\rho^{11}(t)L^\dag+S\rho^{01}(t)L^\dag\xi(t)+L\rho^{10}(t)S^\dag\xi^\ast(t)+S\rho^{00}(t)S^\dag|\xi(t)|^2\right]-\rho^{11}(t)\right\}dN(t),\\
%d\rho^{10}(t)=&\left\{\mathcal{L}^\star_G\rho^{10}(t)+[S\rho^{00}(t),L^\dag]\xi(t)\right\}dt\\
%&+\sqrt{1-r^2}\left[e^{-i\theta}\rho^{10}(t)L^\dag+e^{i\theta}L\rho^{10}(t)+e^{i\theta}S\rho^{00}(t)\xi(t)-K_t\rho^{10}(t)\right]dW(t)\\
%&+\left\{\nu^{-1}_t[L\rho^{10}(t)L^\dag+S\rho^{00}(t)L^\dag\xi(t)]-\rho^{10}(t)\right\}dN(t),\\
%d\rho^{01}(t)=&\left\{\mathcal{L}^\star_G\rho^{01}(t)+[L,\rho^{00}(t)S^\dag]\xi^\ast(t)\right\}dt\\
%&+\sqrt{1-r^2}\left[e^{-i\theta}\rho^{01}(t)L^\dag+e^{i\theta}L\rho^{01}(t)+e^{-i\theta}\rho^{00}(t)S^\dag\xi^\ast(t)-K_t\rho^{01}(t)\right]dW(t)\\
%&+\left\{\nu^{-1}_t[L\rho^{01}(t)L^\dag+L\rho^{00}(t)S^\dag\xi^\ast(t)]-\rho^{01}(t)\right\}dN(t),\\
%d\rho^{00}(t)=&\mathcal{L}^\star_G\rho^{00}(t)dt+\sqrt{1-r^2}\left[e^{-i\theta}\rho^{00}(t)L^\dag+e^{i\theta}L\rho^{00}(t)-K_t\rho^{00}(t)\right]dW(t)\\
%&+\left\{\nu^{-1}_t[L\rho^{00}(t)L^\dag]-\rho^{00}(t)\right\}dN(t).
%\end{aligned}\end{equation}
Here,
\begin{equation}\begin{aligned}
K_t=&e^{-i\theta}\mathrm{Tr}[L^\dag\rho^{11}(t)]+e^{i\theta}\mathrm{Tr}[L\rho^{11}(t)]\\
&+e^{i\theta}\mathrm{Tr}[S\rho^{01}(t)]\xi(t)+e^{-i\theta}\mathrm{Tr}[S^\dag\rho^{10}(t)]\xi^\ast(t),\\
\nu_t=&\mathrm{Tr}[L^{\dag}L\rho^{11}(t)]+\mathrm{Tr}[L^{\dag}S\rho^{01}(t)]\xi(t)\\
&+\mathrm{Tr}[S^{\dag}L\rho^{10}(t)]\xi^\ast(t)+\mathrm{Tr}[\rho^{00}(t)]|\xi(t)|^2,
\end{aligned}\end{equation}
and the initial conditions are $\rho^{11}(0)=\rho^{00}(0)=|\eta\rangle\langle\eta|,~~\rho^{10}(0)=\rho^{01}(0)=0$.
%\begin{equation}\begin{aligned}
%\rho^{11}(0)=\rho^{00}(0)=|\eta\rangle\langle\eta|,~~\rho^{10}(0)=\rho^{01}(0)=0.
%\end{aligned}\end{equation}
\end{corollary}

\subsection{\label{bothhd}Quantum Filter for Both Homodyne Detection Measurements}

\begin{figure}
\centering
\includegraphics[scale=0.3]{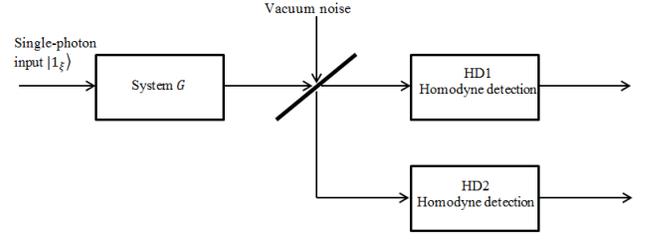}
\caption{Both homodyne detection measurements at the outputs of a beam splitter in quantum system.}
\label{Simultaneoushd}
\end{figure}

In this subsection, we will derive the filter equations for the case of joint homodyne-homodyne measurements, see Fig.~\ref{Simultaneoushd}. Here, by the general measurement equation \eqref{fg}, we choose $F=I,~~G=0$.
%\begin{equation}\nonumber\begin{aligned}
%F=I,~~G=0.
%\end{aligned}\end{equation}
Then, the measurements stochastic equations are given by~\eqref{dy1} and
\begin{equation}\begin{aligned}
dY_{2,t}=&\sqrt{1-r^2}\big(e^{i\theta}dB_{v,t}+e^{-i\theta}dB^\dag_{v,t}\big)\\
&+ir\big\{\big[e^{i\theta}(L+SL_M)-e^{-i\theta}(L^\dag+L^\dag_MS^\dag)\big]dt\\
&+e^{i\theta}SdB_{i,t}-e^{-i\theta}S^{\dag}dB^\dag_{i,t}\big\},
\end{aligned}\end{equation}
where $dY_{2,t}$ is the second channel with homodyne detection measurement. Thus, the corresponding gain $\beta$ can also be calculated by \eqref{beta12}, where $\beta_1$ is given by \eqref{beta1} and $\beta_2$ is given by
\begin{equation}\begin{aligned}
\beta_2=&ire^{i\theta}\tilde{\pi}_t(A\otimes XL+AL_M\otimes XS)\\
&-ire^{-i\theta}\tilde{\pi}_t(A\otimes L^{\dag}X+L^\dag_MA\otimes S^{\dag}X)\\
&-ir\tilde{\pi}_t(A\otimes X)\tilde{\pi}_t[e^{i\theta}(L+SL_M)-e^{-i\theta}(L^\dag+L^\dag_MS^\dag)].
\end{aligned}\end{equation}

Then, in the case of both channels are with homodyne detection measurements, we have the filter equations which are given by the following theorem.

\begin{theorem}\label{theorem2}
Let $\{Y_{i,t},i=1,2\}$ be the two homodyne detection measurements for a quantum system $G$. With single-photon input state, the quantum filter for the conditional expectation in the Heisenberg picture is given by (3.47).
\newcounter{mytempeqncnttt}
\begin{figure*}[!t]
\normalsize
\setcounter{mytempeqncnttt}{\value{equation}}
\setcounter{equation}{46}
\begin{equation}\label{sdehei2}\begin{aligned}
d\pi^{11}_t(X)=&\left\{\pi^{11}_t(\mathcal{L}_GX)+\pi^{01}_t(S^\dag[X,L])\xi^\ast(t)+\pi^{10}_t([L^\dag,X]S)\xi(t)+\pi^{00}_t(S^{\dag}XS-X)|\xi(t)|^2\right\}dt\\
&+\sqrt{1-r^2}\left[e^{i\theta}\pi^{11}_t(XL)+e^{-i\theta}\pi^{11}_t(L^{\dag}X)+e^{-i\theta}\pi^{01}_t(S^{\dag}X)\xi^\ast(t)+e^{i\theta}\pi^{10}_t(XS)\xi(t)-\pi^{11}_t(X)K_{1,t}\right]dW_1(t)\\
&+ir\left[e^{i\theta}\pi^{11}_t(XL)-e^{-i\theta}\pi^{11}_t(L^{\dag}X)-e^{-i\theta}\pi^{01}_t(S^{\dag}X)\xi^\ast(t)+e^{i\theta}\pi^{10}_t(XS)\xi(t)-\pi^{11}_t(X)K_{2,t}\right]dW_2(t),\\
d\pi^{10}_t(X)=&\left\{\pi^{10}_t(\mathcal{L}_GX)+\pi^{00}_t(S^\dag[X,L])\xi^\ast(t)\right\}dt\\
&+\sqrt{1-r^2}\left[e^{i\theta}\pi^{10}_t(XL)+e^{-i\theta}\pi^{10}_t(L^{\dag}X)+e^{-i\theta}\pi^{00}_t(S^{\dag}X)\xi^\ast(t)-\pi^{10}_t(X)K_{1,t}\right]dW_1(t)\\
&+ir\left[e^{i\theta}\pi^{10}_t(XL)-e^{-i\theta}\pi^{10}_t(L^{\dag}X)-e^{-i\theta}\pi^{00}_t(S^{\dag}X)\xi^\ast(t)-\pi^{10}_t(X)K_{2,t}\right]dW_2(t),\\
d\pi^{01}_t(X)=&\left\{\pi^{01}_t(\mathcal{L}_GX)+\pi^{00}_t([L^\dag,X]S)\xi(t)\right\}dt\\
&+\sqrt{1-r^2}\left[e^{i\theta}\pi^{01}_t(XL)+e^{-i\theta}\pi^{01}_t(L^{\dag}X)+e^{i\theta}\pi^{00}_t(XS)\xi(t)-\pi^{01}_t(X)K_{1,t}\right]dW_1(t)\\
&+ir\left[e^{i\theta}\pi^{01}_t(XL)-e^{-i\theta}\pi^{01}_t(L^{\dag}X)+e^{i\theta}\pi^{00}_t(XS)\xi(t)-\pi^{01}_t(X)K_{2,t}\right]dW_2(t),\\
d\pi^{00}_t(X)=&\pi^{00}_t(\mathcal{L}_GX)dt+\sqrt{1-r^2}\left[e^{i\theta}\pi^{00}_t(XL)+e^{-i\theta}\pi^{00}_t(L^{\dag}X)-\pi^{00}_t(X)K_{1,t}\right]dW_1(t)\\
&+ir\left[e^{i\theta}\pi^{00}_t(XL)-e^{-i\theta}\pi^{00}_t(L^{\dag}X)-\pi^{00}_t(X)K_{2,t}\right]dW_2(t).
\end{aligned}\end{equation}
\setcounter{equation}{47}
\hrulefill
\vspace*{4pt}
\end{figure*}
Here,
\begin{equation}\begin{aligned}
K_{1,t}=&e^{i\theta}\pi^{11}_t(L)+e^{-i\theta}\pi^{11}_t(L^\dag)\\
&+e^{-i\theta}\pi^{01}_t(S^\dag)\xi^\ast(t)+e^{i\theta}\pi^{10}_t(S)\xi(t),\\
K_{2,t}=&e^{i\theta}\pi^{11}_t(L)-e^{-i\theta}\pi^{11}_t(L^\dag)\\
&-e^{-i\theta}\pi^{01}_t(S^\dag)\xi^\ast(t)+e^{i\theta}\pi^{10}_t(S)\xi(t),
\end{aligned}\end{equation}
the Wiener processes $W_1(t)$ and $W_2(t)$ are given by
\begin{equation}\begin{aligned}
dW_1(t)=dY_{1,t}-\sqrt{1-r^2}K_{1,t}dt,~~dW_2(t)=dY_{2,t}-irK_{2,t}dt
\end{aligned}\end{equation}
respectively. We have $\pi^{10}_t(X)=\pi^{01}_t(X^\dag)^\dag$, the initial conditions are $\pi^{11}_0(X)=\pi^{00}_0(X)=\langle\eta,X\eta\rangle,~~\pi^{10}_0(X)=\pi^{01}_0(X)=0$.
%\begin{equation}\begin{aligned}
%\pi^{11}_0(X)=\pi^{00}_0(X)=\langle\eta,X\eta\rangle,~~\pi^{10}_0(X)=\pi^{01}_0(X)=0.
%\end{aligned}\end{equation}
\end{theorem}

By the filter equations \eqref{sdehei2} and $\pi^{jk}_t(X)=\mathrm{Tr}[(\rho^{jk}(t))^{\dag}X]$, we also have the quantum filter in the Schr$\ddot{\mathrm{o}}$dinger picture.

\begin{corollary}
With the two homodyne detection measurements, the quantum filter for the system $G$ driven by single-photon input state in the Schr$\ddot{\mathrm{o}}$dinger picture is given by (3.51).
\newcounter{mytempeqncntttt}
\begin{figure*}[!t]
\normalsize
\setcounter{mytempeqncntttt}{\value{equation}}
\setcounter{equation}{50}
\begin{equation}\label{mastereq2}\begin{aligned}
d\rho^{11}(t)=&\left\{\mathcal{L}^\star_G\rho^{11}(t)+[S\rho^{01}(t),L^\dag]\xi(t)+[L,\rho^{10}(t)S^\dag]\xi^\ast(t)+[S\rho^{00}(t)S^\dag-\rho^{00}(t)]|\xi(t)|^2\right\}dt\\
&+\sqrt{1-r^2}\left[e^{-i\theta}\rho^{11}(t)L^\dag+e^{i\theta}L\rho^{11}(t)+e^{i\theta}S\rho^{01}(t)\xi(t)+e^{-i\theta}\rho^{10}(t)S^\dag\xi^\ast(t)-K_{1,t}\rho^{11}(t)\right]dW_1(t)\\
&-ir\left[e^{-i\theta}\rho^{11}(t)L^\dag-e^{i\theta}L\rho^{11}(t)-e^{i\theta}S\rho^{01}(t)\xi(t)+e^{-i\theta}\rho^{10}(t)S^\dag\xi^\ast(t)+K_{2,t}\rho^{11}(t)\right]dW_2(t),\\
d\rho^{10}(t)=&\left\{\mathcal{L}^\star_G\rho^{10}(t)+[S\rho^{00}(t),L^\dag]\xi(t)\right\}dt\\
&+\sqrt{1-r^2}\left[e^{-i\theta}\rho^{10}(t)L^\dag+e^{i\theta}L\rho^{10}(t)+e^{i\theta}S\rho^{00}(t)\xi(t)-K_{1,t}\rho^{10}(t)\right]dW_1(t)\\
&-ir\left[e^{-i\theta}\rho^{10}(t)L^\dag-e^{i\theta}L\rho^{10}(t)-e^{i\theta}S\rho^{00}(t)\xi(t)+K_{2,t}\rho^{10}(t)\right]dW_2(t),\\
d\rho^{01}(t)=&\left\{\mathcal{L}^\star_G\rho^{01}(t)+[L,\rho^{00}(t)S^\dag]\xi^\ast(t)\right\}dt\\
&+\sqrt{1-r^2}\left[e^{-i\theta}\rho^{01}(t)L^\dag+e^{i\theta}L\rho^{01}(t)+e^{-i\theta}\rho^{00}(t)S^\dag\xi^\ast(t)-K_{1,t}\rho^{01}(t)\right]dW_1(t)\\
&-ir\left[e^{-i\theta}\rho^{01}(t)L^\dag-e^{i\theta}L\rho^{01}(t)+e^{-i\theta}\rho^{00}(t)S^\dag\xi^\ast(t)+K_{2,t}\rho^{01}(t)\right]dW_2(t),\\
d\rho^{00}(t)=&\mathcal{L}^\star_G\rho^{00}(t)dt+\sqrt{1-r^2}\left[e^{-i\theta}\rho^{00}(t)L^\dag+e^{i\theta}L\rho^{00}(t)-K_{1,t}\rho^{00}(t)\right]dW_1(t)\\
&-ir\left[e^{-i\theta}\rho^{00}(t)L^\dag-e^{i\theta}L\rho^{00}(t)+K_{2,t}\rho^{00}(t)\right]dW_2(t).
\end{aligned}\end{equation}
\setcounter{equation}{51}
\hrulefill
\vspace*{4pt}
\end{figure*}
Here,
\begin{equation}\begin{aligned}
K_{1,t}=&e^{-i\theta}\mathrm{Tr}[L^\dag\rho^{11}(t)]+e^{i\theta}\mathrm{Tr}[L\rho^{11}(t)]\\
&+e^{i\theta}\mathrm{Tr}[S\rho^{01}(t)]\xi(t)+e^{-i\theta}\mathrm{Tr}[S^\dag\rho^{10}(t)]\xi^\ast(t),\\
K_{2,t}=&e^{-i\theta}\mathrm{Tr}[L^\dag\rho^{11}(t)]-e^{i\theta}\mathrm{Tr}[L\rho^{11}(t)]\\
&-e^{i\theta}\mathrm{Tr}[S\rho^{01}(t)]\xi(t)+e^{-i\theta}\mathrm{Tr}[S^\dag\rho^{10}(t)]\xi^\ast(t),
\end{aligned}\end{equation}
and the initial conditions are $\rho^{11}(0)=\rho^{00}(0)=|\eta\rangle\langle\eta|,~~\rho^{10}(0)=\rho^{01}(0)=0$.
%\begin{equation}\begin{aligned}
%\rho^{11}(0)=\rho^{00}(0)=|\eta\rangle\langle\eta|,~~\rho^{10}(0)=\rho^{01}(0)=0.
%\end{aligned}\end{equation}
\end{corollary}

\subsection{Simulation Results} \label{susec:simulation}

Here we apply the filter equations derived in subsection~\ref{bothhd} to the problem of exciting a two-level atom with a continuous-mode single-photon, \cite{GOUGH12QUANTUM}. This system can be parameterized as follows. The scattering is $S=I$, the coupling operator is $L=\kappa\sigma_-$ with coupling strength $\kappa=1$. The atom is taken to be in the ground state initially $|g\rangle\langle g|$ with the Hamiltonian $H=0$. The wave packet $\xi(t)$ for the single-photon is given by
\begin{equation}\begin{aligned}
\xi(t)=\left(\frac{\Omega^2}{2\pi}\right)^{1/4}\exp\left[-\frac{\Omega^2}{4}(t-t_0)^2\right],
\end{aligned}\end{equation}
where $t_0$ is the peak arrival time and $\Omega$ is the frequency bandwidth of the wave packet.

Now we choose $\Omega=1.46$ and wish to calculate the exciting probability for the atom as a function of time.
%
%In what follows, we denote the exciting probability for master equations by
%\begin{equation}\begin{aligned}
%P_e(t)=\mathrm{Tr}[\hat{\rho}^{11}(t)|e\rangle\langle e|]
%\end{aligned}\end{equation}
%where $\hat{\rho}^{11}(t)$ is the solution to \eqref{mastereq} and
The exciting probability for quantum filtering equations is given by
\begin{equation}\label{58}\begin{aligned}
P^c_e(t)=\mathrm{Tr}[\rho^{11}(t)|e\rangle\langle e|]
\end{aligned}\end{equation}
where $\rho^{11}(t)$ is the solution to \eqref{mastereq2}.

\begin{figure}
\centering
\includegraphics[scale=0.24]{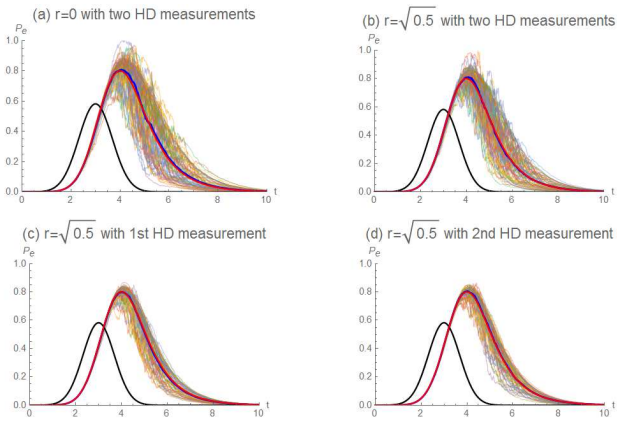}
\caption{(Color online) The exciting probability for a two-level system interacting with one photon in a Gaussian pulse shape with different beam splitter parameters. The black line is the wave packet $|\xi(t)|^2$, the red line is $P_e(t)$ given by the master equation, the colorful lines are the trajectories $P^c_e(t)$ and the blue line denotes the average of these trajectories.}
\label{Simulation}
\end{figure}

In Fig.~\ref{Simulation}, 72 different stochastic trajectories are simulated as colorful lines in each case given by \eqref{58}. Fig.~\ref{Simulation}(a) ($r=0$) denotes the ideal case which is equivalent to the single measurement (HD1) without any noise, \cite{GOUGH12QUANTUM}. For $r=1$, the case will be similar to Fig.~\ref{Simulation}(a) since the single measurement becomes HD2. We can see that many of the stochastic trajectories begin to decay after the bulk of the wave packet, i.e., $t=4$. Meanwhile, some trajectories continue to rise towards $P^c_e(t)=1$, it means that the atom may be fully excited. In Fig.~\ref{Simulation}(b),  $r=\sqrt{0.5}$, that is the output field is contaminated by vacuum noise. Nevertheless, it can be seen that by means of joint measurement the estimation performance is close to those for the ideal case. The exciting probabilities become bad if we only use single measurement, see Fig.~\ref{Simulation}(c) and (d). By comparing Fig.~\ref{Simulation}(b), (c) and (d), it is clear that multiple measurements is much better.

\section{CONCLUSIONS}\label{sec:conc}

In this paper, we have derived the quantum filter for a quantum system driven by single-photon input state with multiple compatible measurements. Particularly, the explicit form of stochastic master equations with two homodyne detection measurements and a combination of homodyne detection and photon-counting are given. A numerical study of a two-level system driven by a single-photon state demonstrated the advantage of filtering design based on multiple measurement when the output filed is contaminated by quantum vacuum noise.

In our full paper preparation, we are considering the stability of single photon filtering in this paper. Our approach is based on applying the method applied in~\cite{amini2014stability}. As a further direction, we can study the filtering problem when we consider the multi-photon input state~\cite{SONG13MULTI}. Also, we may take into account imperfections in measurements. Moreover, showing the stability of multi-photon filtering is in the perspective of our research.

%\section*{ACKNOWLEDGMENT}


\begin{thebibliography}{99}

\bibitem{amini2014stability}
H. Amini, C. Pellegrini, and P. Rouchon, Stability of continuous-time quantum filters with measurement imperfections, Russ. J. Math. Phys., vol. 21, no. 3, pp. 297-315, 2014.

\bibitem{amini2013feedback}
H. Amini, R. A. Somaraju, I. Dotsenko, C. Sayrin, M. Mirrahimi, and P. Rouchon, Feedback stabilization of discrete-time quantum systems subject to non-demolition measurements with imperfections and delays, Automatica, vol. 49, no. 9, pp. 2683-2692, 2013.

\bibitem{belavkin1989nondemolition}
V. P. Belavkin, Nondemolition measurements, nonlinear filtering and dynamic programming of quantum stochastic processes, Modeling and Control of Systems, pp. 245-265, Springer, 1989.

\bibitem{belavkin1995quantum}
V. P. Belavkin, Quantum filtering of Markov signals with white quantum noise, Quantum Communications and Measurement, pp. 381-391, Springer, 1995.

\bibitem{carmichael2009open}
H. Carmichael, An open systems approach to quantum optics: lectures presented at the Universit{\'e} Libre de Bruxelles, vol. 18, Springer Science \& Business Media, 2009.

\bibitem{carvalho2012cavity}
A. R. R. Carvalho, M. R. Hush, and M. R. James, Cavity driven by a single photon: Conditional dynamics and nonlinear phase shift, Phys. Rev. A, vol. 86, no. 2, pp. 023806, 2012.

\bibitem{chia2011quantum}
A. Chia and H. M. Wiseman, Quantum theory of multiple-input--multiple-output Markovian feedback with diffusive measurements, Phys. Rev. A, vol. 84, no. 1, pp. 012120, 2011.

\bibitem{dum1992monte}
R. Dum, A. S. Parkins, P. Zoller, and C. W. Gardiner, Monte Carlo simulation of master equations in quantum optics for vacuum, thermal, and squeezed reservoirs, Phys. Rev. A, vol. 46, no. 7, pp. 4382, 1992.

\bibitem{eichler2011experimental}
C. Eichler, D. Bozyigit, C. Lang, L. Steffen, J. Fink, and A. Wallraff, Experimental state tomography of itinerant single microwave photons, Phys. Rev. Lett., vol. 106, no. 22, pp. 220503, 2011.

\bibitem{woolley2015quantum}
M. F. Emzir, M. J. Woolley, and I. R. Petersen, Quantum filtering for multiple diffusive and Poissonian measurements, J. Phys. A: Math. Theor., vol. 48, no.38, pp. 385302-385319, 2015.

\bibitem{gardiner2004quantum}
C. W. Gardiner and P. Zoller, Quantum noise, Springer, 2004.

\bibitem{GOUGH09SERIES}
J. E. Gough and M. R. James, The series product and its application to quantum feedforward and feedback networks, IEEE Trans. Autom. Control, vol. 54, pp. 2530-2544, 2009.

\bibitem{GOUGH12QUANTUM}
J. E. Gough, M. R. James, H. I. Nurdin, and J. Combes, Quantum filtering for systems driven by fields in single-photon states or superposition of coherent states, Phys. Rev. A, vol. 86, pp. 043819, 2012.

\bibitem{mckeever2004deterministic}
J. McKeever, A. Boca, A. D. Boozer, R. Miller, J. R. Buck, A. Kuzmich, and H. J. Kimble, Deterministic generation of single photons from one atom trapped in a cavity, Science, vol. 303, no. 5666, pp. 1992-1994, 2004.

\bibitem{Hendra14QUANTUM}
H. I. Nurdin, Quantum filtering for multiple input multiple output systems driven by arbitrary zero-Mean jointly Gaussian input fields, Russ. J. Math. Phys., vol. 21, no. 3, pp. 386-398, 2014.

\bibitem{Sayrin2011}
C. Sayrin, I. Dotsenko, X. Zhou, B. Peaudecerf, T. Rybarczyk, S. Gleyzes, P. Rouchon, M. Mirrahimi, H. Amini, M. Brune,
J. M. Raimond, and S. Haroche, Real-time quantum feedback prepares and stabilizes photon number states,
Nature, vol. 477, no. 7362, pp. 73-77, 2011.

\bibitem{SONG13MULTI}
H. Song, G. Zhang, and Z. Xi, Continuous-mode multi-photon filtering, arXiv preprint: 1307.7367, 2013.

\bibitem{wang2011efficient}
Y. Wang, J. Min{\'a}{\v{r}}, L. Sheridan, and V. Scarani, Efficient excitation of a two-level atom by a single photon in a propagating mode, Phys. Rev. A, vol. 83, no. 6, pp. 063842, 2011.

\bibitem{wiseman2009quantum}
H. M. Wiseman and G. J. Milburn, Quantum measurement and control, Cambridge University Press, 2009.

%\addtolength{\textheight}{-12cm} % This command does not take effect until the next page

\bibitem{yuan2002electrically}
Z. Yuan, B. E. Kardynal, R. M. Stevenson, A. J. Shields, C. J. Lobo, K. Cooper, N. S. Beattie, D. A. Ritchie, and M. Pepper, Electrically driven single-photon source, Science, vol. 295, no. 5552, pp. 102-105, 2002.

\bibitem{ZJ12}
G. Zhang and M. R. James, Quantum feedback networks and control: a brief survey, Chin. Sci. Bull., vol. 57, no. 18, pp.2200-2214, 2012.

\end{thebibliography}
\end{document}